\documentclass[10pt,journal,compsoc]{IEEEtran}

\ifCLASSOPTIONcompsoc
  \usepackage[nocompress]{cite}
\else
  \usepackage{cite}
\fi
\ifCLASSINFOpdf
   \usepackage[pdftex]{graphicx}
  \DeclareGraphicsExtensions{.pdf,.jpeg,.png}
\else
\fi
\usepackage{amsmath}
\usepackage{amssymb}
\usepackage{algpseudocode,algorithm}
  \usepackage[caption=false,font=footnotesize,labelfont=sf,textfont=sf]{subfig}
\newcommand\MYhyperrefoptions{bookmarks=true,bookmarksnumbered=true,
pdfpagemode={UseOutlines},plainpages=false,pdfpagelabels=true,
colorlinks=true,linkcolor={black},citecolor={black},urlcolor={black},
pdftitle={Integration of CUDA Processing within HPX},
pdfsubject={Paper},
pdfauthor={P. Diehl et al.},
pdfkeywords={Asynchronous many task systems, CUDA, parallelism, concurrency, HPX}}
\ifCLASSINFOpdf
\usepackage[\MYhyperrefoptions,pdftex]{hyperref}
\else
\usepackage[\MYhyperrefoptions,breaklinks=true,dvips]{hyperref}
\usepackage{breakurl}
\fi
\usepackage{xcolor}
\usepackage{booktabs}
\usepackage{listings}
\definecolor{darkgreen}{rgb}{0.1, 0.7, 0.3}
\newcommand{\I}[1]{\textit{#1}}

\usepackage{todonotes}

\usepackage{pgfplots}
\pgfplotsset{compat=1.14}

\usepackage{pdfpages}

\begin{document}
%
\title{Integration of CUDA Processing within the C++ library for parallelism and concurrency (HPX)}
%
%
%
%

\author{Patrick Diehl, Madhavan Seshadri, Thomas Heller, Hartmut Kaiser
\IEEEcompsocitemizethanks{\IEEEcompsocthanksitem P. Diehl, T. Heller, and H. Kaiser were with the Center for Computation and Technology, Louisiana State University, LA,
US.\protect\\
E-mail: P. Diehl see https://orcid.org/0000-0003-3922-8419
\IEEEcompsocthanksitem M. Seshadri was with the Nanyang Technological University, Singapore
\IEEEcompsocthanksitem T. Heller was with the Department of Computer Science, Friedrich-Alexander-University of Erlangen-Nürnberg, Germany
\IEEEcompsocthanksitem H. Kaiser was with the Department of Computer Science, Louisiana State University, LA. USA.
\IEEEcompsocthanksitem P. Diehl, T. Heller, and H. Kaiser were at the Ste$\vert\vert$ar group}
}

\IEEEtitleabstractindextext{%
\begin{abstract}
Experience shows that on today's high performance systems the utilization of
different acceleration cards in conjunction with a high utilization of all other
parts of the system is difficult. Future architectures, like exascale clusters, are expected to aggravate
this issue as the number of cores are expected to increase and memory hierarchies
are expected to become deeper. One big aspect for distributed applications is
to guarantee high utilization of all available resources, including local or remote acceleration cards
on a cluster while fully using all the available CPU resources and the
integration of the GPU work into the overall programming model.

For the integration of CUDA code we extended HPX, a general
purpose C++ run time system for parallel and distributed applications of
any scale, and enabled asynchronous data transfers from and to the GPU device
and the asynchronous invocation of CUDA kernels on this data.
Both operations are well integrated into the general programming model of HPX
which allows to seamlessly overlap any GPU operation with work on the main
cores. Any user defined CUDA kernel can be launched on any (local
or remote) GPU device available to the distributed application.

We present asynchronous implementations for the data transfers and
kernel launches for CUDA code as part of a HPX asynchronous execution
graph. Using this approach we can combine all remotely and locally available
acceleration cards on a cluster to utilize its full performance capabilities.
Overhead measurements show, that the integration of the asynchronous operations
(data transfer + launches of the kernels) as part of the HPX execution graph
imposes no additional computational overhead and significantly eases
orchestrating coordinated and concurrent work on the main cores and the used GPU devices.
\end{abstract}

}

\maketitle

\IEEEdisplaynontitleabstractindextext

%
\IEEEpeerreviewmaketitle

\section{Introduction}
The biggest disruption in the path to exascale will occur at the
intra-node level, due to severe memory and power constraints per
core, many-fold increase in the degree of intra-node parallelism,
and to the vast degrees of performance and functional heterogeneity
across cores. The significant increase in complexity
of new platforms due to energy constraints, increasing parallelism
and major changes to processor and memory architecture,
requires advanced programming techniques that are portable across
multiple future generations of machines~\cite{XStackFOA}. This trend has already
manifested itself for some time in the domain of accelerator and co-processor
boards.

It is well known that a large part of the available compute power of a machine (in terms of FLOPS) today is
provided by various accelerators and co-processors. Especially general purpose GPUs however
require special programming languages and techniques. Unfortunately, those devices are architecturally not too well
integrated with the main cores. This requires special effort from the programmer in terms of
managing a heterogeneous code-base, having to explicitly manage the data transfer to and
from the devices and the execution of special kernels. In order for this scheme to be scaleable,
special care is required to
a)~keep the main cores busy while kernels are being executed on a GPU,
and b)~hide the latencies and overheads of data transfers behind useful work. In short, the currently
available solutions make it very hard to achieve \I{scalability}, \I{programmability}, and \I{performance portability} for applications running on heterogeneous resources.

In this paper we will focus on a technique and programming environment, which overcomes part of the above mentioned problems by transparently enabling asynchronous data transfer and kernel execution for CUDA,
while still being able to concurrently execute tasks on the main cores in a seamless way. All GPU operations are represented as asynchronous tasks similar to any other parallel task run on a main core. This facilitates an easy way
to express dependencies which is a critical precondition for managing parallel execution graphs in the HPX framework~\cite{Heller2017}.

The presented solution not only transparently facilitates the hiding of latencies of data transfers to and
from accelerators and the asynchronous execution of compute kernels, it also provides a framework for
load balancing work across the system over
large amount of (possibly remote) accelerator cards. 
For the kernels themselves, the solution still relies on
the proven CUDA technology of the existing and widely used programming environments for GPUs. We do however expose the events generated by CUDA as C++ \I{future} objects
to the user (see Section~\ref{sec:futurization}) which enables
to integrate the data transfer and the execution of the kernels with the overall parallel execution flow on the
main cores.

This paper makes the following contributions to the C++ library for parallelism and concurrency:
\begin{itemize}
\item HPXCL provides an API for transparently enabling asynchronous data transfer and kernel execution for CUDA,
\item all API functions return a C++ \I{future} object, which can be used within the synchronization mechanics provided by HPX for the integration in its asynchronous execution graph,  
\item the CUDA specific elements, \emph{e.g.}\ blockSize, threadSize, and the kernel code, are not hidden from the user which allows the easy integration of existing CUDA code into the HPX framework.
\end{itemize}

The remainder of this paper is structured as follows: In Section~\ref{sec:related:work} the related work is presented. In Section~\ref{sec:implementation} HPX's details, which are used for the integration, are briefly introduced. In Section~\ref{sec:design} examples for accessing all remote and local CUDA devices and the complete work flow for data transfers and launching a kernel are shown. Section~\ref{sec:overhead} shows the overhead measurements compared to a native CUDA implementation. Finally, Section~\ref{sec:conclusion} concludes this presented approach.
\section{Related work}
\label{sec:related:work}
This section provides a brief overview of related approaches for the integration of CUDA processing. For unifying the data transfer between different compute nodes via the Message Passing Interface (MPI), Nvidia provides CUDA-aware MPI~\cite{cudampi}. Here, Unified Virtual Addressing (UVA) is utilized to combine the host memory and device memory of a single node into one virtual address space. Thus, pointers to data on the device can be handled by MPI directly and can be integrated in the MPI message transfer pipeline. However, the synchronization of kernel launches and data transfer is not addressed here.

The Chapel programming language~\cite{Chamberlain:2007:PPC:1286120.1286123} provides \emph{parallel} and \emph{distributed} language features for parallel and distributed computing. Thus, there is a distinct separation of parallelism and locality in this programming model. In addition, the specific functionality of the acceleration card is hidden from the user trough parallel feature of the language.

HPX.Compute~\cite{heller2016closing} is fully compatible to the C++ $17$ standard N$4578$~\cite{Committee4578} and implements the triple define execution model: \emph{targets}, \emph{allocator} for memory allocation purposes, and \emph{executors} for specifying when, where, and how the work is executed. HPX.Compute supports CUDA devices. 

HPX.Compute SYCL~\cite{Copik:2017:USI:3078155.3078187} provides a new back end for HPX.Compute utilizing SYCL, a Khronos standard for single-source programming of OpenCL devices.

Kokkos~\cite{CarterEdwards20143202} provides abstractions for parallel code execution and data management for CUDA, OpenMP, and Pthreads via a C++ library. Similar to CUDA-aware MPI, it does support a specific memory space (CudaVMSpace) and allocation within this space are accessible from host and device. The computational bodies (kernels) are passed to Kokkos via function objects or lambda function to the parallel executor.

The Phalanx programming model~\cite{Garland:2012:DUP:2388996.2389087} provides a unified programming model for heterogeneous machines on a single node. For multiple nodes of a distributed-memory cluster the GASNet~\cite{bonachea2017gasnet} run time is utilized. This model provides its interface of generic and templated functions via C++ template library. The implementation of the kernel function is identical for the CUDA and OpenMP backend by providing an abstraction level and the asynchronous launches of the kernels are synchronized using events. 

RAJA~\cite{hornung2014raja} introduces its fundamental concept for separating the loop body from the loop transversal by introducing the \emph{forall} feature where a sequential code block is defined. By providing the execution policy and a \emph{indexset} the separation is modeled. Its concept is used in physics code, where plenty of for loops are used to do multi physics simulations.

Thrust~\cite{BELL2012359} is based on the Standard Template Library (STL) and CUDA::thrust is an extension for providing CUDA devices and provides a set of common parallel algorithms. CUDA::thrust tries to hide CUDA specific language features by providing a similar API to the C++ API.

The programming language X$10$~\cite{ebcioglu2004x10,ebcioglu2005x10} provides an open-source tool chain, which translates the X$10$ code to C++. As the underlying programming model APGAS is used on the GPU and the CPU. Thus, the CUDA threads are defined as APGAS activities by specifying a first loop over the CUDA blocks and a second loop over the CUDA threads. Within the second for loop the sequential code for each CUDA thread is defined. The philosophy here was to use existing X$10$ constructs to hide the CUDA specific semantics. 

Table~\ref{tab:sum:app} summarizes all these different approaches. Two different type of approaches can be seen in the related work. First, the approaches of providing a programming language are different from the presented approach, since the C++ library for parallelism (HPX) is extended for the integration of CUDA.  Second, the library-based approaches are more similar to HPXCL, but all these approaches try to hide the CUDA specific language features as much as possible from the user. HPXCL instead allows the user to define CUDA specific feature, \emph{e.g.}\ provide kernel functions and specify block and thread sizes. Therefore, HPXCL focuses on the approach to integrate existing CUDA kernels in the asynchronous execution graph of HPX. For hiding CUDA specific language features the HPX.Compute framework is more suitable.

\begin{table}[tb]
\caption{Summary of the different approaches for the integration of CUDA. The second column lists the technologies provided by each approach, the third column indicates the type of the approach, and the last column provides the reference.}
\begin{tabular}{lllc} \toprule
Name & Technology & Type & Ref \\ \midrule
Chapel & CUDA,Xeon Phi  & Language & \cite{Chamberlain:2007:PPC:1286120.1286123}\\
CUDA-aware MPI & MPI,CUDA & Lib  & \cite{cudampi} \\
HPX.Compute & HPX,CUDA & C++ Lib & \cite{heller2016closing}   \\
HPX.Compute SYCL & HPX,OpenCL & C++ Lib & \cite{Copik:2017:USI:3078155.3078187} \\
Kokkos & CUDA,OpenMP & C++ Lib & \cite{CarterEdwards20143202} \\
Phalanx & CUDA,OpenMP, GASNet & C++ Lib & \cite{Garland:2012:DUP:2388996.2389087} \\
RAJA & OpenMP,CUDA & C++ Lib & \cite{hornung2014raja}  \\
Thrust & CUDA,OpenMP & C++ Lib & \cite{BELL2012359} \\
X10 & CUDA& Language & \cite{ebcioglu2004x10} \\\bottomrule
\end{tabular}
\label{tab:sum:app}
\end{table}

\section{HPX's basics}
\label{sec:implementation}
\begin{figure}[tb]
\centering
\includegraphics[width=0.65\linewidth]{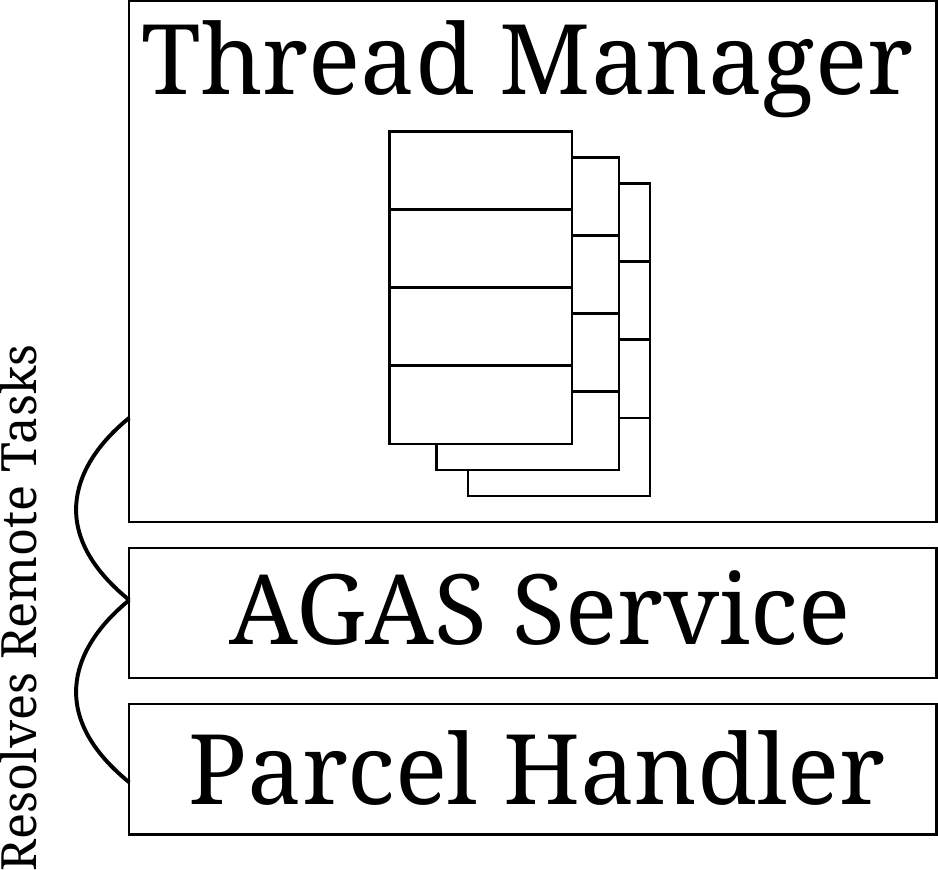}
\caption{Run time components of HPX (Thread manager,Active Global Address Space (AGAS) Service, and Parcel Handler), which resolve the remote tasks. The thread manager deals with the light-weight user level threads and provides a high-level API.The Active Global Address Space (AGAS) Service provides Global IDs to each allocated objects for hiding the explicit message passing. The parcel handler provides the communication between different nodes via remote procedure call / Active Messaging. Adapted from~\cite{Heller2017}.}
\label{fig:main:components}
\end{figure}

The HPX compute language (HPXCL)~\cite{Diehl2018} is an extension for the Open Source C++ library for parallelism and concurrency (HPX)~\cite{Heller2017}. The asynchronous many tasks (AMT) programming paradigm provided by HPX is extended with the asynchronously data transfer from the host to device (and vice versa) and asynchronously kernel launches. The synchronization between the tasks on a CUDA devices and CPUs is realized by the concept of futurization. We briefly review the main components of HPX, which are utilized for the integration of asynchronous tasks and synchronization. For more details about HPX we refer to~\cite{tabbal2011preliminary,kaiser2014hpx,Heller2017}. Figure~\ref{fig:main:components} shows the three main components of HPX, which resolve the remote tasks. 

-- The \emph{Thread Manager}~\cite{kaiser2009parallex} deals with the light-weight user level threads and provides a high level API. Within HPX pre-defined scheduling policies are defined: 1) \emph{static} means that one queue is attached to one core, 2) \emph{thread local} which is HPX's default scheduling policy and means one queue is attached to one core, but in addition to the static scheduling policy, task stealing from neighboring cores on the same node is enabled and 3) \emph{hierarchical} means that there is a tree of queues and new tasks are attached to the root of the tree and move down when a core fetches new work. There is always the possibility to define application specific scheduling policies, but HPXCL uses the \emph{static} one. 

-- \emph{Active Global Address Space (AGAS) Service}~\cite{kaiser2014hpx} supports the distributed computing. Each object within AGAS is represented by its Global ID (GID) to hide explicit message passing. Thus, its address is not bound to a specific locality on the system and its remote or local access is unified. This feature allows us to provide  the same API for a local or remote CUDA device in our environment.

-- \emph{Parcel Service}~\cite{ac:2017,kaiser2009parallex}
For the communication between different nodes in the cluster environment, HPX is utilization remote procedure call (RPC) /    Active Messaging. In the terminology of HPX an active message is a so-called Parcel which provides calls to remote nodes in a C++ fashion. For the communication between nodes either the tcp protocol or the Message passing Interface (MPI) is used.
\subsection{Futurization}
\label{sec:futurization}
The API exposed by HPXCL is fully asynchronous and returns a \lstinline!hpx::future!.
The \emph{future} is an important building block in the HPX infrastructure. By providing an
uniform asynchronous return value, percolation as discussed in this paper is tightly integrated with
any other application written in HPX and allows writing \emph{futurized} code by employing the
standard-conforming API functions such as \lstinline!hpx::future<T>::then! and
\lstinline!hpx::when_all<T>! for composition, and \lstinline!hpx::dataflow! to build implicit parallel execution
flow graphs~\cite{Kaiser:2015:HPL:2832241.2832244}. This allows for a unified continuation based
programming style for both, regular host code and device code, and is an excellent tool to close the
architectural gap between classic GPUs and CPUs.

\section{Design of the implementation}
\label{sec:design}
\begin{figure}[tb]
\begin{center}
  \includegraphics[width=\linewidth]{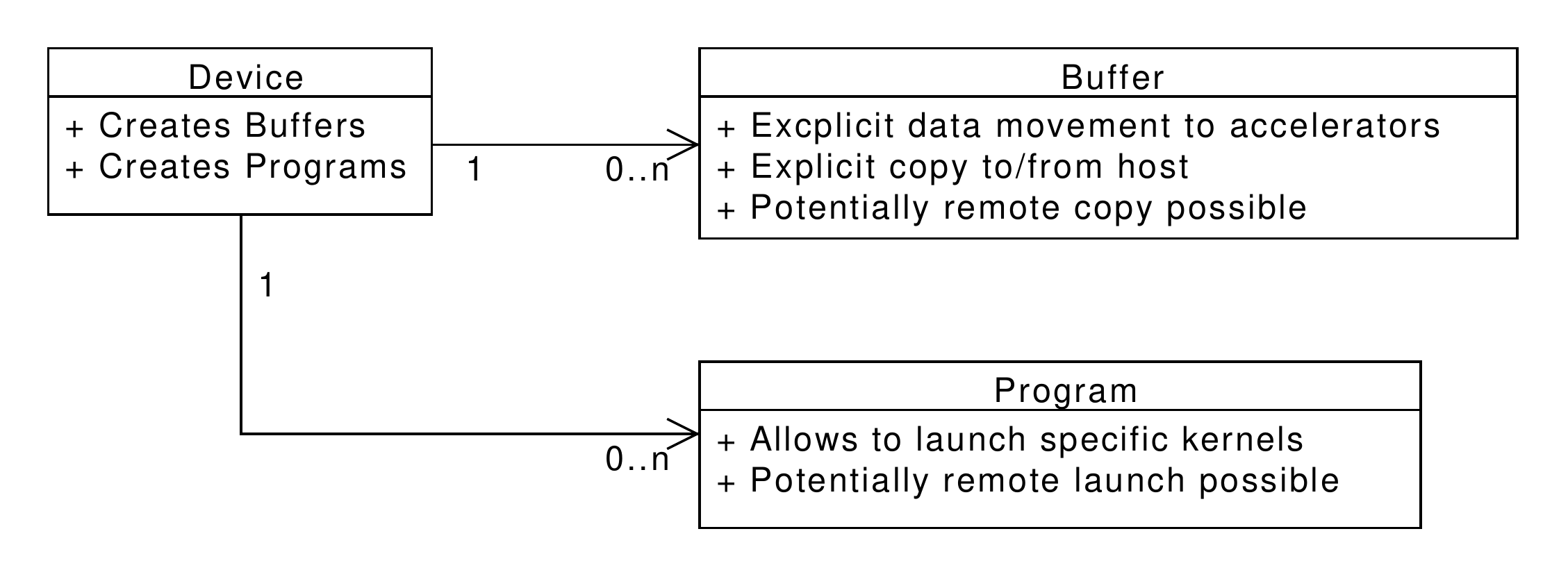}
  \caption{Class diagram of  Buffer, Device and Program and the functionality provided by each class. The device represents a logical device on a remote or local locality. A program which handles the compilation and the potential remote launches of kernels. The memory of a device is represented by a buffer. Adapted from~\cite{Diehl:2017}.}
  \label{fig:hpxcl-class_diagram}
\end{center}
\end{figure}

Figure~\ref{fig:hpxcl-class_diagram} shows the basic structure and relation between the classes. The user facing API is represented by client side objects referencing the object in AGAS via its Global ID (GID). This approach offers two benefits:
a) Both copying and passing the object to different localities are now
transparent in regard to the object pointing to either a remote or local
object.
b) They are also transparent in regard to where the actual representation of
the object resides, i.e. is the data on the locality where it is needed.

For example, if a buffer or kernel is created for a specific device, the client
side object only references the actual memory. Once those objects are used, the operation is completely local to the accelerator and the associated kernel is executed where the data lives. The device code to be executed is compiled just-in-time, that is, each accelerator action is available in source code or compatible
binary form and can be sent and executed on the respective device, after it has been compiled for a
specific device.

This represents an implementation of \emph{percolation}, which allows data and code
to be freely moved around in the (possibly) distributed system using the parcel service. The data is moved between the node using either tcp or MPI. The functions exposed by this API are inherently asynchronous (\emph{i.e.}\ return a \emph{future} object representing its return value) and therefore
allows to naturally overlap unavoidable latencies when compiling kernels, copying data, or executing
device functions. Each of these functions is attached to an light-weight user level thread using the \emph{static} scheduling policy.

-- A \emph{device} is the logical representation of an accelerator and defines the functionality to
execute kernels, create memory buffers, and to perform synchronization. HPXCL exposes functionality to discover local
and remote devices within a system. Each device is equipped with its own, platform dependent
asynchronous work queue, while still allowing cross-device synchronization through HPX's \emph{future}
API.

-- A \emph{buffer} represents memory which is allocated on a specific \emph{device}. The operations
defined on a buffer are related to copying data from and to the host system and between different
devices. While the content itself is not directly addressable through AGAS, the asynchronous copy functionality
is which allows effective memory exchange between different entities of a possible distributed
memory system. The copy functions return futures which can be used as dependencies to either kernel
calls and additionally allows for naturally overlapping of communication with computation.

-- A \emph{program} serves as the compiled form of code that is to be executed on a specific device.
By executing kernels, buffers need to be supplied as arguments. Executing a kernel returns a \I{future} object as well.
Those futures can also be used to express data flow dependencies from memory copy operations or other
previous calls to kernels.

\subsection{Access of local and remote devices}
Listing~\ref{lst:devices} shows the one line of source code to discover local and remote devices within a system. The method \lstinline|get_all_devices| takes two arguments (the major and minor compute capability) and returns a \lstinline|std::vector| with all available devices \lstinline|hpx::cuda::device| having at least a compute capability as specified. The method returns a future and therefore \lstinline|.get()| has to be called to receive the content of this future. Note that all device objects have the same API independent of the device is a remote or a local device.
\begin{lstlisting}[language=C++,caption=Gathering all remote and local CUDA devices on the cluster having CUDA compute capability of at least $1.0$. Note that the function call returns a \emph{future} and calling the \lstinline|.get()| function return the content of this future.,label={lst:devices},float=tb]
int hpx_main(int argc, char* argv[]) {

//Get list of available CUDA Devices.
std::vector<hpx::cuda::device> devices = 
 	hpx::cuda::get_all_devices(1,0).get();
 	
return hpx::finalize();
}
\end{lstlisting}
\subsection{Workflow of HPXCL}
Listing~\ref{lst:all} shows the work flow for running a CUDA kernel for computing the sum of $n$ elements and stores the result in one variable. In Line~\ref{code:1} all available devices within the cluster environment are collected. In Line~\ref{code:2}--\ref{code:3} the data on the host is allocated. In Line~\ref{code:4} the first device in the list which can be either a remote or local device is selected to run the computation. In Line~\ref{code:5}--\ref{code:6} three buffers are generated which means that internally a \lstinline|cudaMalloc| is called. After the creation of each buffer the data is copied into this buffer which means internally a \lstinline|cudaMemcpyAsync| is done and the future of this function calls are stored in a vector \lstinline|futures| for later synchronization. In Line~\ref{code:7} the CUDA kernel is loaded from the file \emph{kernel.cu}, the run time compilation of the kernel is started using NVRTC - CUDA Runtime Compilation, and the future is added to the vector of futures. In Line~\ref{code:8}--\ref{code:9} the configuration of the kernel launch is defined. In Line~\ref{code:10}--\ref{code:11} the arguments of the CUDA kernel are defined. In Line~\ref{code:12} the compilation of kernel and the copy from the data to the CUDA device have to be finished to execute the kernel properly. Therefore, a barrier with \lstinline|hpx::wait_all| is introduced to assure that all these dependencies are finished. In Line~\ref{code:13} the kernel is executed and finally in Line~\ref{code:14} the result of the execution is copied to the host using \lstinline|cudaMemcpyAsync|.\\

Note that we use native CUDA functionality to synchronize the asynchronous CUDA function calls, but we hide this from the user by returning a future object. This allows the users to combine these tasks with tasks on the CPU within the cluster environment in a unified fashion. In addition, the usage of remote or local devices has the same unified API and HPXCL internally copy the data to the node where the data is needed.

\begin{lstlisting}[language=C++,caption=Workflow of a simple application.,label={lst:all},float=*ptb,numbers=left,stepnumber=1,escapechar=|]
	// Get list of available Cuda Devices.
	std::vector<device> devices = get_all_devices(2, 0).get(); |\label{code:1}|
	// Allocate the host data
	unsigned int* input; |\label{code:2}|
	unsigned int* n;
	unsigned int* res;
	cudaMallocHost((void**)&input, sizeof(unsigned int)* 1000);
	cudaMallocHost((void**)&result,sizeof(unsigned int));
	cudaMallocHost((void**)&n,sizeof(unsigned int));
	memset (input,1,1000);	
	result[0] = 0;
	n[0] = 1000; |\label{code:3}|
	// Create a device component from the first device found
	device cudaDevice = devices[0]; |\label{code:4}|	
	// Create a buffers and copy data into them
	std::vector<hpx::lcos::future<void>> futures;
	buffer outbuffer = cudaDevice.create_buffer(SIZE * sizeof(unsigned int)).get(); |\label{code:5}|	
	futures.push_back(outbuffer.enqueue_write(0, SIZE * sizeof(unsigned int), inputData));
	buffer resbuffer = cudaDevice.create_buffer(sizeof(unsigned int)).get();
	futures.push_back(resbuffer.enqueue_write(0,sizeof(unsigned int), result));
	buffer lengthbuffer = cudaDevice.create_buffer(sizeof(unsigned int)).get();
	futures.push_back(lengthbuffer.enqueue_write(0,sizeof(unsigned int), n)); |\label{code:6}|	
	// Compile the CUDA Kernel
	program prog = cudaDevice.create_program_with_file("kernel.cu").get();
	futures.push_back(prog.build("sum")); |\label{code:7}|	
	// Prepare the configuration of the kernel
	hpx::cuda::server::program::Dim3 grid; |\label{code:8}|	
	hpx::cuda::server::program::Dim3 block;
	grid.x = grid.y = grid.z = 1;
	block.x = 32;
	block.y = block.z = 1; |\label{code:9}|	
	// Set the arguments of the kernel
	std::vector<hpx::cuda::buffer>args; |\label{code:10}|	
	args.push_back(outbuffer);
	args.push_back(resbuffer);
	args.push_back(lengthbuffer); |\label{code:11}|	
	// Synchronize the copy of data to the device and the compilation of the kernel
	hpx::wait_all(data_futures); |\label{code:12}|	
	//Run the kernel at the default stream
	prog.run(args,"sum",grid,block).get(); |\label{code:13}|	
	//Copy the result back
	unsigned int* res = resbuffer.enqueue_read_sync<unsigned int>(0,sizeof(unsigned int)); |\label{code:14}|	
	
\end{lstlisting}
\section{Overhead measurements}
\label{sec:overhead}
The unified API and utilizing HPX as an additional layer introduces some overhead. To measure the  which is introduced by additional layer of HPX, the same benchmark is done by using native CUDA. Therefore, the native CUDA functions call used in each method and there synchronization were analyzed and were re-implemented without using the additional layer of HPX. For all benchmarks the same kernel, block size, and thread size were used. Note that for these measurements, the focus is on the overhead introduced by HPXCL and not on the optimization of the
kernels for obtaining the optimal performance. The authors are aware that several highly optimized benchmark suites are available. However, comparing against these suites would not measure the overhead introduced by HPXCL, since the memory bandwidth and the computational throughput are measure for a kernel~\cite{cudametrics}. Since we use the same kernel, the device performance would not change significantly and instead the overall end to end performance can be compared.

In the Appendix: the software, the hardware, the operating system and drivers, and compilers are listed in detail, to enhance the reader's ability to easily reproduce these benchmarks on their own cluster environment. In order to alleviate start-up times, we repeated the algorithm for 11 iterations and took the mean execution time out of the last ten iterations, the first iteration was considered to be the warm-up (meant to be ignored). For the corresponding details, we refer to the appendix.

\subsection{Single device}
For the single devices overhead measurements, a Nvidia Tesla K$40$ and Nvida Tesla K$80$ in to different compute nodes (bahram,reno) were utilized. For each benchmark the identical kernel was executed using the native CUDA implementation and the equivalent HPX implementation.   
\subsubsection{Stencil kernel (Intel Parallel Research Kernels)}
\label{sec:bench:stencil}
\begin{figure}[tb]
\centering
\begin{tikzpicture}
\begin{axis}[
  ylabel={Execution time (ms)},
  xlabel={Length of the vector $n$},
  legend style={at={(0.5,-0.2)},anchor=north},
  legend columns=2,
  grid=both]
  \addplot[black,mark=*] table [x, y, col sep=comma] {stencilCudaK40.dat};
  \addplot[black,mark=square*] table [x, y, col sep=comma] {stencilHPX1K40.dat};
  \addplot[blue,mark=*] table [x, y, col sep=comma] {stencilCudaK80.dat};
  \addplot[blue,mark=square*] table [x, y, col sep=comma] {stencilHPX1K80.dat};
  \legend{CUDA K$40$\\HPX K$40$\\CUDA K$80$\\HPX K$80$\\}
\end{axis}
\end{tikzpicture}
\caption{Comparison for the overhead introduced by the additional layer of HPXCL for the stencil benchmark. The measurements were performed on a single Tesla K$40$ and a single K$80$. The native CUDA implementation is compared against the HPX implementation using one CPU.}
\label{fig:bench:stencil}
\end{figure}
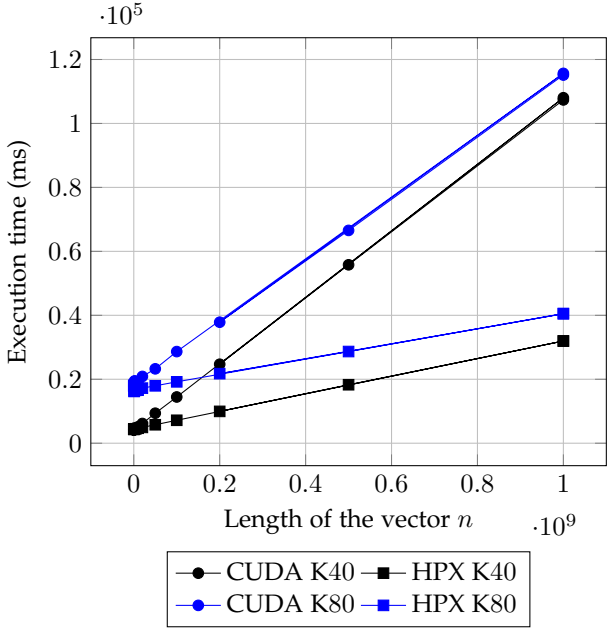

This benchmark, where a $3$-point stencil $s(x_i):=0.5x_{i-1}+x_i+0.5x_{i+1}$ is computed for a vector $X:=\lbrace x_i,\ldots, X_n \vert x_i \in \mathbb{R} \rbrace$ has been defined within the Intel Parallel Research Kernels (PRK)~\cite{van2014parallel} developed by Intel labs. The block size was one and the thread size was $32$ for all benchmarks. The aim of this benchmark is to test the computation and synchronization of the application. Figure~\ref{fig:bench:stencil} shows the execution time vs. length of the vector for the K$40$ (black lines) and for the K$80$ (blue lines). In both cases, the HPX implementation is $\approx 28\%$ faster than the native CUDA implementation and the trend of the two lines is nearly linear. Thus, overlapping computation and data transfer utilizing futures has helped reduce the overall compute time. Note, that in this benchmark the CUDA code was executed sequentially and the HPX code was using the asynchronous functionality within the CUDA SDK. In the next benchmark, the asynchronous functionality will be used in the native CUDA implementation as well.
\subsubsection{Partition example}
\label{sec:bench:partition}
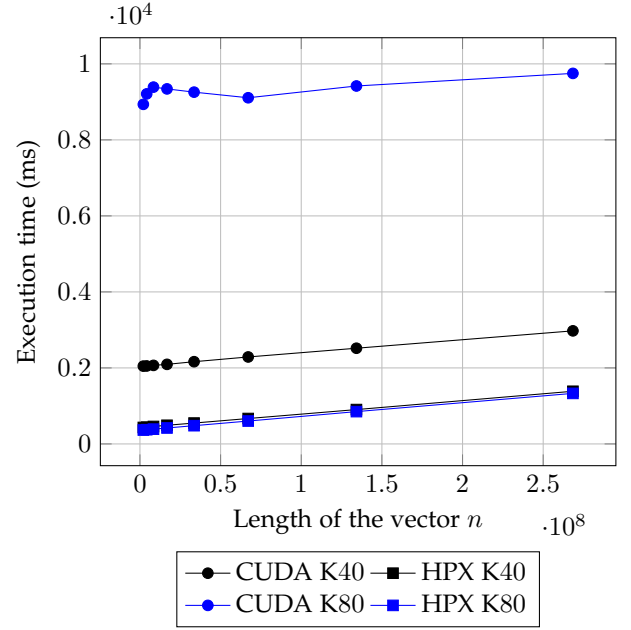
\begin{figure}[tb]
\centering
\begin{tikzpicture}
\begin{axis}[
  ylabel={Execution time (ms)},
  xlabel={Length of the vector $n$},
  legend style={at={(0.5,-0.2)},anchor=north},
  legend columns=2,
  grid=both]
  \addplot[black,mark=*] table [x, y, col sep=comma] {partitionCudaK40.dat};
  \addplot[black,mark=square*] table [x, y, col sep=comma] {partitionHPXK40.dat};
  \addplot[blue,mark=*] table [x, y, col sep=comma] {partitionCudaK80.dat};
  \addplot[blue,mark=square*] table [x, y, col sep=comma] {partitionHPXK80.dat};
  \legend{CUDA K$40$\\HPX K$40$\\CUDA K$80$\\HPX K$80$\\}
\end{axis}
\end{tikzpicture}
\caption{Comparison for the overhead introduced by the additional layer of HPXCL for the partition benchmark. The measurements were performed on a single Tesla K$40$ and a single K$80$. The native CUDA implementation using asynchronous function calls and synchronization utilizing streams is compared against the HPX implementation using one CPU.}
\label{fig:bench:partiton}
\end{figure}

This benchmark focuses on the asynchronous data transfer and the efficient overlapping between communication and computation. The native CUDA implementation was adopted from~\cite{async_cuda} and asynchronous function calls are used in both implementations. In this benchmark, a kernel $k(x_i):=\sqrt{\sin^2{i}+\cos^2(i)}$ is computed for a vector $X:=\lbrace x_i,\ldots, X_n \vert x_i \in \mathbb{R} \rbrace$. The length of the vector is given by $n= 2^m * 1024 * blockSize * p$, where $m=\lbrace 1,2,\ldots,7,8\rbrace$, $blockSize=256$, and $p=4$ is the amount of partitions. The vector is divided in $p$ partitions and each partition is asynchronously copied to the CUDA device, the kernel $k$ is executed, and the result is asynchronously copied back to the host, see Algorithm~\ref{benchmark-partition}. CUDA streams are used for the synchronization for the native CUDA implementation and the HPX implementation.
\begin{algorithm}[h]
\begin{algorithmic}
\State Init $X$
\For{ $i=0$ ; $i <$ $p$ ; ++$i$ }
\State cudaMemcpyAsync($X_{i}$,cudaMemcpyHostToDevice)
\EndFor
\For{ $i=0$ ; $i <$ $p$ ; ++$i$ }
\State Apply kernel $k$ to partition $X_i$
\EndFor
\For{ $i=0$ ; $i <$ $p$ ; ++$i$ }
\State cudaMemcpyAsync($R_{i}$,cudaMemcpyDeviceToHost)
\EndFor
\end{algorithmic}
\caption{Multiple Partitions Benchmark}
\label{benchmark-partition}
\end{algorithm}
Figure~\ref{fig:bench:partiton} shows the execution time vs. length of the vector for the K$40$ (black lines) and for the K$80$ (blue lines) using four partitions. In both cases, the HPX implementation is $\approx 4\%$ faster than the native CUDA implementation and the trend of the two lines is nearly linear. It is clearly shown that the speed-up of HPX is reduced by a factor of $\approx 4$  when the native CUDA implementation utilizes asynchronous function calls. This benchmarks shows that the overhead introduced by HPXCL is negligible, even when the CUDA kernel is compiled at run time, for large enough vector sizes. 

\subsubsection{Mandelbrot example using concurrency with CPUS}
\label{sec:bench:mandel}
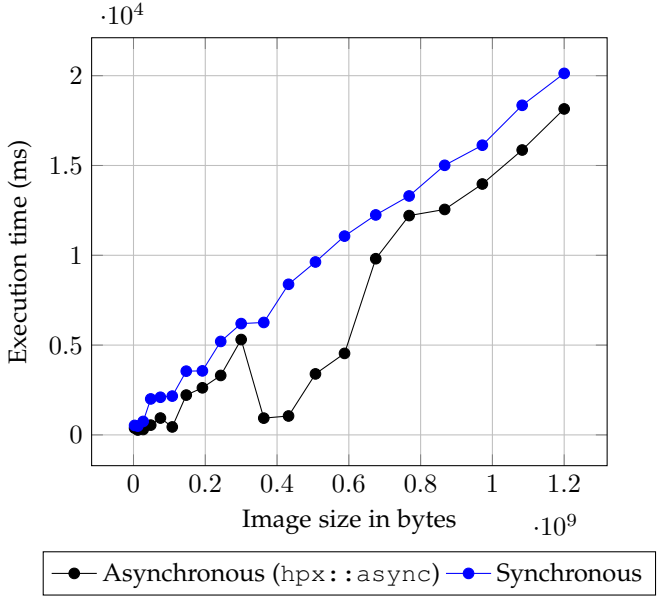
\begin{figure}[tb]
\centering
\begin{tikzpicture}
\begin{axis}[
  ylabel={Execution time (ms)},
  xlabel={Image size in bytes},
  legend style={at={(0.5,-0.2)},anchor=north},
  legend columns=2,
  grid=both]
  \addplot[black,mark=*] table [x, y, col sep=comma] {imageasync.dat};
  \addplot[blue,mark=*] table [x, y, col sep=comma] {imagesync.dat};
  \legend{Asynchronous (\lstinline|hpx::async|)\\Synchronous\\}
\end{axis}
\end{tikzpicture}
\caption{Comparison of synchronous and asynchronous writing the resulting image of the Mandelbrot set to the file system. Both measurements were done on one K$80$ card using HPXCL.}
\label{fig:bench:mandelbrot}
\end{figure}

The Mandelbrot set is a set of complex numbers for which $c$ does not diverge from $0$ for the function $f_c(z)=z^2+c$ when integrated from zero. The pixels are then colored based on how rapidly the function value diverges from zero. In this example the Mandelbrot set for increasing images sizes is computed on a $K80$ GPU using HPXCL and saved to the file system as a PNG image. Figure~\ref{fig:bench:mandelbrot} shows the computation time (blue line) when the image is synchronously written after the computation. The black line shows the computational time when the concurrency with the CPU of the HPXCL framework is used and the image is written asynchronously using \lstinline|hpx::async| to the file system. Using the concurrency with the CPU decreases the computational time and this feature is beneficial for example to write results to the file system and asynchronously compute the next iteration.

\subsection{Multiple devices}
For the multiple device partition benchmark two Nvidia Tesla K$80$ cards containing two sub cards in one node (bahram) were utilized. Note, that each K$80$ has a dual-GPU design and therefore, $2\times2$ GPUs are available.
\subsubsection{Partition example}
\label{sec:bench:partitionm}
\begin{figure}[tb]
\centering
\begin{tikzpicture}
\begin{axis}[
  ylabel={Execution time (ms)},
  xlabel={Length of the vector $n$},
  legend style={at={(0.5,-0.2)},anchor=north},
  legend columns=2,
  grid=both]
  \addplot[black,mark=*] table [x, y, col sep=comma] {partion_k80_1_average.dat};
  \addplot[blue,mark=*] table [x, y, col sep=comma] {partion_k80_1_average_hpx.dat};
  \addplot[black,mark=square*] table [x, y, col sep=comma] {partion_k80_2_average.dat};
  \addplot[blue,mark=square*] table [x, y, col sep=comma] {partion_k80_2_average_hpx.dat};
  \addplot[black,mark=diamond*] table [x, y, col sep=comma] {partion_k80_3_average.dat};
  \addplot[blue,mark=diamond*] table [x, y, col sep=comma] {partion_k80_3_average_hpx.dat};
  \addplot[black,mark=+] table [x, y, col sep=comma] {partion_k80_4_average.dat};
  \addplot[blue,mark=+] table [x, y, col sep=comma] {partion_k80_4_average_hpx.dat};
  \legend{CUDA $1\times$K$80$\\HPX $1\times$K$80$\\CUDA $2\times$K$80$\\HPX $2\times$K$80$\\CUDA $3\times$K$80$\\HPX $3\times$K$80$\\CUDA $4\times$K$80$\\HPX $4\times$K$80$\\}
\end{axis}
\end{tikzpicture}
\caption{Comparison for the overhead introduced by the additional layer of HPXCL for the partition benchmark on multiple devices. The vector is sliced in $1,2,3,$ and $4$ partitions and each partition is handled by one of the K$80$ cards. }
\label{fig:bench:partiton:multi}
\end{figure}
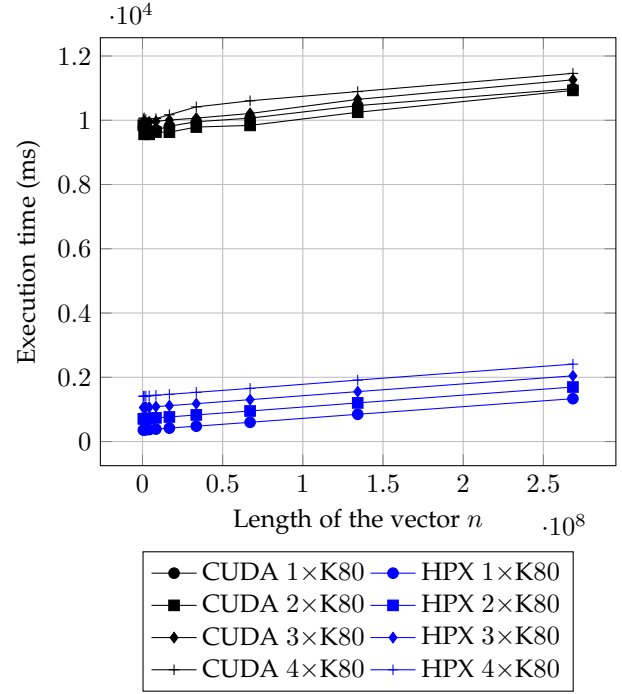

The benchmark describes in Section~\ref{sec:bench:partition} is modified, such that each partition of the vector is handled by one of the Nvidia Tesla K$80$ cards in the bahram node. The vector with the length $n= 2^m * 1024 * blockSize$, where $m=\lbrace 1,2,\ldots,7,8\rbrace$ and $blockSize$ is $256$, is sliced in $1,2,3,$ and $4$ partitions and each partition is handled by one of the K$80$ cards. The black lines show the execution time for the native CUDA implementation and the blue lines for $1$ up to $4$ K$80$ devices, see Figure~\ref{fig:bench:partiton:multi}. For the native CUDA implementation a increase of computation time is seen when going from one physical card to two physical cards. Once the dual-GPU architecture is used, a increase in computational time is seen. For the HPXCL implementation the same behavior for the computational time (blue lines) is seen. The difference between the computational times is one order of magnitude. First, the exact behavior for the dual-GPU case was obtained. This could be improved by using new CUDA features, but since the focus is on the overhead and not the performance this is not relevant for this paper. Second, also for the Multiple GPUs case, the overhead introduced by HPXCL is small and the execution time is faster.

\section{Conclusion and future work}
\label{sec:conclusion}
In this paper we present an abstraction over CUDA -- HPXCL -- which is tightly integrated into the HPX general purpose parallel run time system. This allows seamless integration into a fully heterogeneous application accessing local and remote devices in a unified fashion. Note that within this implementation the CUDA specific functionality, \emph{e.g.}\ blockSize, threadSize, and kernel code, is not hidden from the application developer. Therefore, HPXCL is suitable to integrate existing CUDA kernel into the asynchronous execution of HPX.

Our overhead evaluation showed that the performance cost of
using such an abstraction is minimal and our implementation could outperform the native CUDA implementation. For the sake of fairness, the same features, \emph{e.g.}\ CUDA streams, asynchronous memory functionality (\lstinline|cudaMemcpyAsync|), and \lstinline|cudaStreamSynchronize| for synchronization, were used in both implementations. However, HPX uses light-weighted threads and even with the usage of one CPU, more than one light-weighted thread is generated. Thus, HPX has the benefit of using more threads which results in the faster computational times. Also, the CUDA kernel is compiled at run time using NVRTC - CUDA Runtime Compilation for HPXCL and compiled at compile time for the native CUDA application. Within this set up a fair comparison of the execution times is hard to archive.

Nevertheless, the presented abstraction can be considered to improve programmability and maintainability of heterogeneous, distributed workloads. The data transfers and the launch of the kernel can be easily integrated in the asynchronous workload on the CPU, like the writing of the image in the Mandelbrot benchmark while the next image size is computed. 

This work showed the proof of concept for an unified API for the integration CUDA to HPX by introducing a negligible overhead. A next step would be to do some performance benchmarks against existing benchmark suits which would require more optimization in the naive CUDA kernels. Two possible future directions for HPXCL are PeridynamicHPX~\cite{2018arXiv180606917D} and a simple computational fluid dynamics (CFD)\footnote{\url{https://github.com/ltroska/nast_hpx}} solver for in compressible Navier-Stokes equations~\cite{Trosk2016}. 

For most CFD problems, e.q. driven cavity, natural convection and the K\'{a}rm\'{a}n vortex street, the matrix-vector operation is the bottle neck for the overall computational time. Here, the benefits of GPUs for solving such problems is easily integrated in the existing HPX code with the HPXCL library. For PeridynamicHPX, a non-local fracture mechanics code, one large portion of the computational costs is the neighbor search. Here, this task could be done on GPUs, where fast algorithms for this task are available.

\section*{Acknowledgements}
This material is based upon work supported by the NSF Award $1737785$ and a Google Summer of Code stipend.

\ifCLASSOPTIONcompsoc
\else
\fi

\ifCLASSOPTIONcaptionsoff
  \newpage
\fi



\bibliographystyle{IEEEtran}
\bibliography{IEEEabrv,Bibliography}
%





\end{document}